May 20th, 2016

# Bubbles: a data management approach to create an advanced industrial interoperability layer for critical systems development applying reuse techniques[1]

Aleksander Lodwich[1] and Jose María Alvarez-Rodríguez

(1) ITK Engineering AG
Liebknechtstr. 33
D-70565 Stuttgart-Vaihingen, Germany
Phone Number: +49 711 933157-393
{aleksander.lodwich}@itk-engineering.de

(2) Department of Computer Science
Carlos III University of Madrid,
Avd. De la Universidad, 20, Leganés, Madrid, Spain, 28911
Phone number: +34 916249115
{josemaria.alvarez}@uc3m.es

**Abstract.** The development of critical systems is becoming more and more complex. The overall tendency is that development costs raise. In order to cut cost of development, companies are forced to build systems from proven components and larger new systems from smaller older ones. Respective reuse activities involve good number of people, tools and processes along different stages of the development lifecycle which involve large numbers of tools. Some development is directly planned for reuse. Planned reuse implies excellent knowledge management and firm governance of reusable items. According to the current state of the art, there are still practical problems in the two fields, mainly because the governance and knowledge management is fragmented over the tools of the toolchain. In our experience, the practical effect of this fragmentation is that involved ancestor and derivation relationships are often undocumented or not exploitable. Additionally, useful reuse is almost always dealing with heterogeneous content which must be transferred from older to newer development environments. In this process, interoperability proves either as biggest obstacle or greatest help. In this paper, authors connect the topics interoperability and knowledge management and propose to seek for ubiquitous reuse via advanced interoperability features. A single concept from a larger Technical Interoperability Concept (TIC) with the name *bubble* is presented. Bubbles are expected to overcome existing barriers to cost-efficient reuse in systems and software development lifecycles. That is why, the present paper introduces and defines bubbles by showing how they simplify application of repairs and changes and hence contribute to expansion of reuse at reduced cost.

## Introduction

The governance of reused design is among the toughest questions in systems development. It becomes more pressing due to raising individualization of products which must be delivered at cost of mass production (Benavides, Segura, and Ruiz-Cortés 2010). A recent survey study with practitioners revealed that insufficient reuse is the second most ranking problem experienced by developers with MBSE practice (Marko et al. 2014). For this reason

1 *This is the extended version of the paper to ease the review in case of acceptance the paper will be reduced to meet the length extension of the conference.*

it appears justified to investigate new ways to produce reuse-friendly and cost-friendly development technology for the industry.

However, after a long time, reuse promises (Jacobson, Griss, and Jonsson 1997) (Karlsson 1995) are still far from reaching the major objective of optimizing system development efforts, even though tools, techniques, methods and languages and the overall understanding of a software-based system (aka cyber-physical system) have dramatically changed since the NATO Software Engineering Conference in 1968 (Mcilroy 1969).

What is reuse? In the field of software engineering, reuse is commonly defined as a process to systematically specify, produce, classify, retrieve and adapt software artifacts for the purpose of using them in a development process. This simple and powerful definition was introduced (Mili, Mili, and Mili 1995) four decades ago with the aim of overcoming the problem of software failures. In general, software reuse (Krueger 1992) may have the potential of increasing productivity of software developers, improve software quality and create a cost efficient development environment. However, both technical and non-technical issues for a limited systems and software reuse can be found (Smolárová and Návrat 1997): 1) economical, organizational, educational or psychological issues and 2) lack of standards to represent software artifacts, and lack of reusable component libraries or appropriate tools for boosting reuse and interoperability among tools. This situation is becoming critical in the development of critical systems in which it is not just a matter of reusing software artifacts but any artifact generated during the development lifecycle such as a document or even a process.

Building on existing technical issues, systems and software engineering techniques have been widely studied (Boehm 1981) (Kim and Stohr 1998) (Mili 2002) to support the classical principles of software reuse (Krueger 1992) (Biggerstaff and Richter 1989) : abstraction, selection, specialization and integration. More specifically, abstraction (i.e. management of the intellectual complexity of a software artifact) can be considered the essential feature for any reuse technique to specify when an artifact could be reused and how to reuse it. Selection refers to the discovery of software artifacts, covering from the representation and storage to the classification (Prieto-Díaz 1991), location and comparison. Specialization consists on the set of parameters and transformations required to reuse a software artifact, while integration refers to the capability of software systems to communicate, collaborate and exchange data.

Thus, the reusability factor of any artifact will directly depend on how they are abstractly described, how they can be selected and specialized for reuse, and how they will integrate in the new complete system. Furthermore, a reuse approach implies that every artifact generated during the development lifecycle is not any more an isolated requirement specification, model, piece of source code or test case, but a knowledge and organizational asset.

On the other hand, knowledge management techniques (Nonaka and Takeuchi 1995) have gained enough momentum in the systems engineering area to elevate the meaning of the implicit knowledge (Jose Maria Alvarez-Rodríguez et al. 2015) encoded into requirements, models or source code. The development of a system can be seen as an underlying multimode (different types of nodes) and multilayered (baselines, processes or structures) graph, an industrial knowledge graph that compiles all people, tools and artifacts generated during the development life-cycle. However, already the mere transition from one development branch to another or the impact of fixing a bug can imply dramatically changes in the development process. Although system development methodologies and processes are perfectly defined, it is still not clear how to pack specific parts to enable further reuse and to ease system engineers' tasks such as automatic configuration of development environments.

In this sense, knowledge management techniques (Nonaka and Takeuchi 1995) can be applied to capture, structure, store and disseminate system artifacts to directly support the aforementioned reuse principles of selection and integration. In order to ease the reuse of system artifacts, it is required to select an adequate knowledge representation paradigm (Davis, Shrobe, and Szolovits 1993) (Groza et al. 2009). After a long time (Hull and King 1987), this problem still persists since a suitable representation format (and syntax) can already be reached in several ways. But, on the other hand, knowledge management also implies the standardization of data and information exchange to support other application services such as business analytics or knowledge discovery.

In this context, the current trend in systems and software development lies in applying an interoperability-based approach (instead of software product line or product line engineering approaches) to enable tools and, by extension, engineers to collaborate and share information under a common data model that can be accessed using standard communication mechanisms (HTTP-based services). The Open Service for Lifecycle Collaboration (OSLC) is becoming a reality to build a collaborative system development ecosystem (Manikas and Hansen 2013) for product and application developments (Thüm et al. 2014) through the definition of data shapes that serve as a contract to get access to information resources through REST (Representational State Transfer) services.

Although OSLC can solve the problem of interoperability (syntax and semantics of data items and communication protocols) through a set of standardized services, there is a gap between the sharing and representation of information and exploitation of the underlying knowledge. Some approaches are emerging to tackle this issue by providing mechanisms to index and retrieve any kind of artifact but the problem of providing configurable and flexible development environments still persists (ubiquitous reuse of information and processes). Furthermore, the use of web-based techniques to access information resources is sometimes misleading (e.g. link stability) since the web and, more specifically, web-based services, were not conceived to address the typical issues in a systems development environment such as configuration management (INCOSE 2004). Thus, although knowledge management techniques are perceived by engineers as necessary they also imply a lot of extra effort and, so far, they can only be exploited for some specific purposes.

Some problems can be solved easier if there is a central authority storing all design data in a single service which is responsible for keeping all the information related to each other in the right way. However, more and more industrial actors are concerned with distributed services and their inability to have a single source of data. This forces the industry to look for technology dealing well with distributed data. For instance, in the context of the CRYSTAL project[2], a large scale research project, main activity is to investigate potentials and limits of OSLC in industrial real-life use cases. Although OSLC is a quite good approach for sharing information resources, OSLC-based services live as isolated applications that can collaborate together only in a limited way when avoiding a better organization of a hyperlink-based environment. This limited scope of responsibility in OSLC has impact on reuse. Successful reuse strategies often strongly rely on specific features of every service, tool or data storage. Such features would be described as “conglomeration”, “replication”, “transformation”, “relocation”,” relinking” and “reconciliation of dependencies”. Because they are missing these operations are performed by developers. In practical life, the manual act of reuse is performed by groups of professionals with better or lesser discipline for applying certain rules.

IBM’s report on strategic reuse patterns in product line engineering (Eran and Scouler 2014) is representative of current understanding of the problem addressed in this paper. The basic concepts of the approach are: baselines, streams and configurations (see Figure 1).

2 http://www.crystal-artemis.eu/

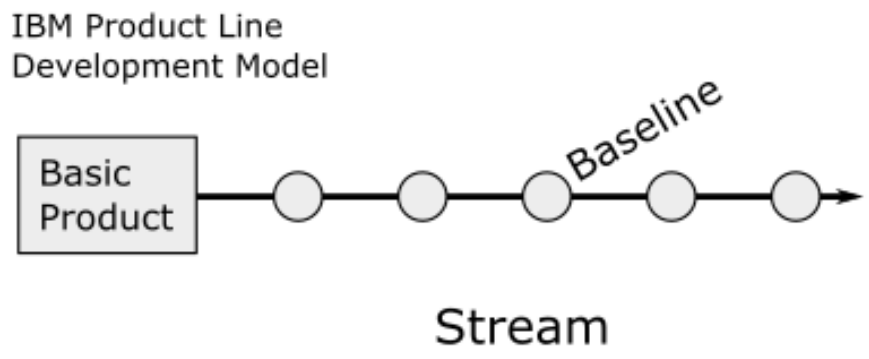

**Figure 1: Basic concepts of IBM's strategic reuse concept for product lines.**

Eran's and Scouler's argument for a strategically introducible three level reuse model is built on a very popular model to describe development of related product designs: the *product lines*. A key concept to the product lines is the *stream* as has been shown in Figure 1. Under ideal circumstances the stream represents a linear history of design development fixated in baselines. A baseline is an immutable snapshot of data. In more generic sense, a baseline is an immutable set of entities and their relationships.

The advantage of baselines is lying in their immutability and a known "address" (often a human readable name). A user of a baseline can rely on a safe ground for his work. A baseline can also be used to start new work or to revert subsequent erroneous edits. For this to be possible a baseline must represent a known quality standard. In that sense, baselines have an elevated status compared with other snapshots. Baselines are often only committed after a quality assurance process has taken place. Baseline management defines quality standards and processes for shooting baselines.

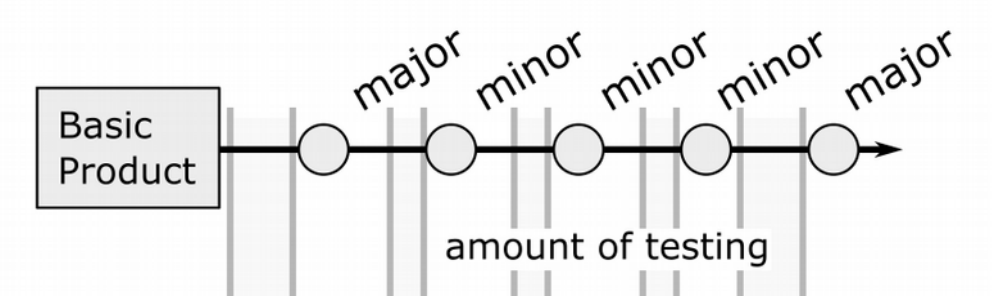

**Figure 2: Baselines can have different state or level attributed to them.**

Baseline management can be organized around various quality levels. Baselines with very high quality standards can endure longer periods of time. Baselines with lower quality levels will endure only short time before they are surpassed by newer baselines of same level. At the lowest quality level is the single snapshot which might not test quality of committed data at all. This arrangement can be used to decouple slow and fast development processes from interacting counterproductively. Slow activities can rely on a stable ground when fast processes keep committing edits. How engineering departments use their baselining features may also depend on external reasons like the duty to document progress.

Reusing design in this context will mean to define a baseline to be reused and to create a new stream from it (see Figure 3). At this point, the design of the stream is intended to diverge from other existing streams. By doing so, a large body of reviewed content is inherited. Very challenging in this context is selective branching of streams or merging of selected features from various streams because baselines cannot be subdivided a posteriori into more reusable entities. The definitions and processes to branch a stream of baselines into new streams can be called stream management. Frequently, it comes along with changes to content of variant management databases, streams must get new names, new staff must be allocated to its development, etc.

One of the extinguished challenges in reusing design via streams is the problem of how problems and bug fixes are recorded and deployed. Because the fundamental unit is a baseline, which is by its nature static, it is not possible to edit a baseline that "caused" the problem. As long as some baseline is truly correct it is not a problem to branch from it. This is the "fun" part which is shown on the left side of the branch event in Figure 3. Change requests occurring after the branching event are the “pain” part because for every ticket regarding any of the reused elements a) the proper locations must be identified and b) individually applied.

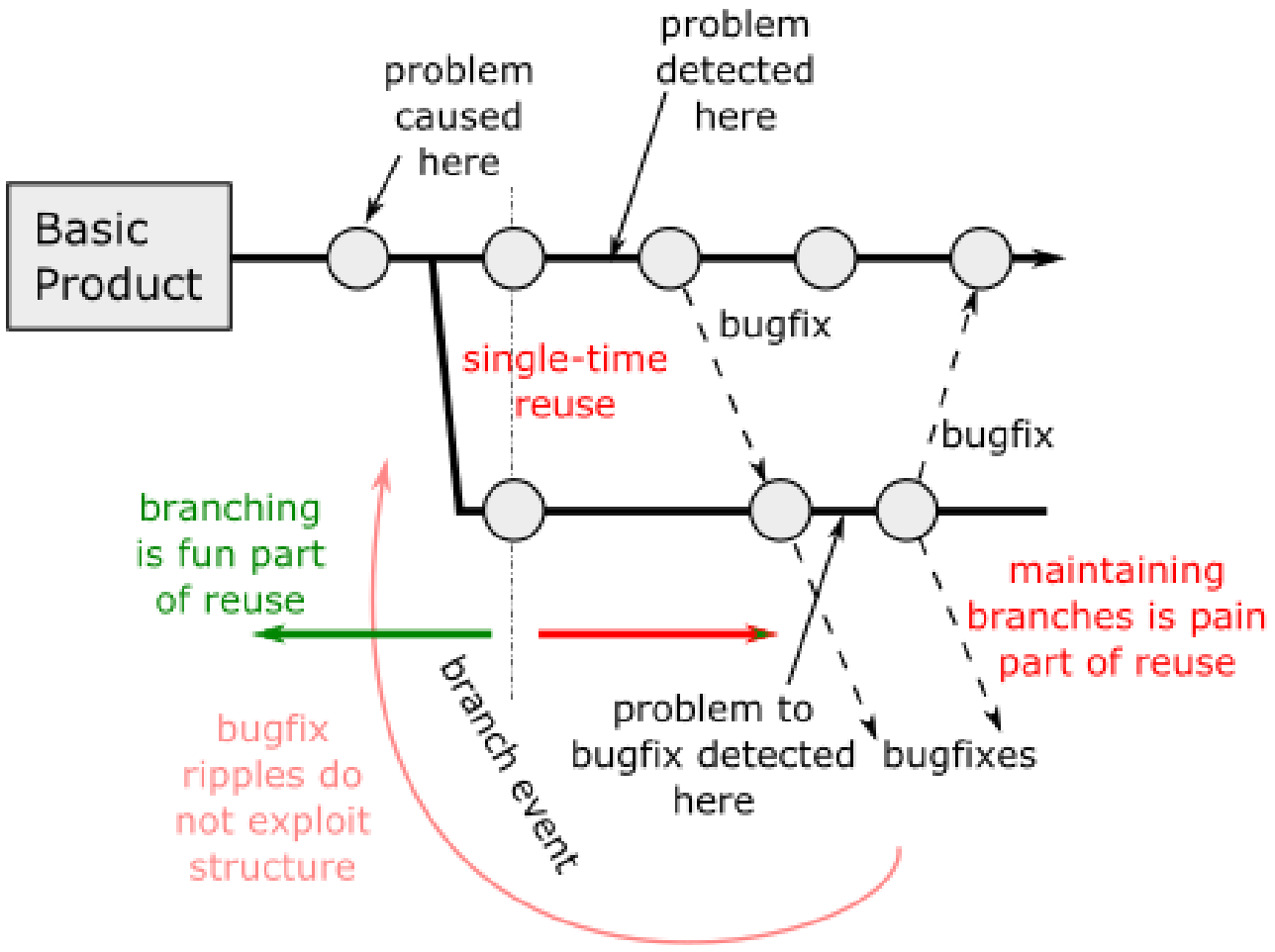

**Figure 3: It is possible to branch baselines but not to edit them. Fixing a problem in a design forces the product line owner to fix the problems separately for each stream head.**

A typical systems project will rely on several major design databases with baselining capabilities. Top level stream management is not automatically applied to participating databases despite this could be considered best practice. If you split a stream of product design into two this should be reflected by respective branches in all participating design storages / databases. Even if streams are aligned then application of changes to a stream are far from pleasant activity (cf. red b’s in Figure 4).

Variants shall be briefly discussed in this context which are often understood as projections of an over-unity design. Creation of configurable over-unity designs is basically assuming that features are orthogonal and can be activated or deactivated. But the problem of difficult to apply repairs and changes also applies to over-unity designs. In fact, over-unity designs add challenges because variants have the same potential to diverge as designs modeled as streams. Over time configurable designs show accumulation of exceptions which require multiple fixing places in the over-unity design in excess of existing in multiple streams. It would be an improvement if design could be managed in such a way that choice between over-unity approaches and streams was less final and application of changes more holistic. This consideration has led to certain bubble features which can be only mentioned here.

In general tools vendors assume that problems associated with applying changes are minor. Advanced platforms like the IBM Rational suite provides change set management which can be pushed forward for their application. However, the arguments here are often linear. In more realistic cases bug hopping can be observed even in sophisticated environments. That is because a stream of configurations is not relying on any kind of structural continuity as can be seen in Figure 5. It shows a stream landscape that is slowly diverging. Motivations behind the diversions are twofold: In anticipation of future change requests and problem reports every engineering department is attempting to keep the number of individual streams as low as possible. If a stream would branch and not require a change then teams prefer not to introduce a new stream. Such a shared stream holds the promise to reduce work related to applying changes but even after short time the teams are overwhelmed with complexity to understand shared design features. Moreover, customers require updates to otherwise discontinued elements or a team will explore alternative implementations. This process will create new streams not directly motivated by top level stream management. As long as those baselines are not in an active configuration by which they could receive bug reports and change sets they will not care to apply them. However, nothing prevents that they become part of a configuration later.

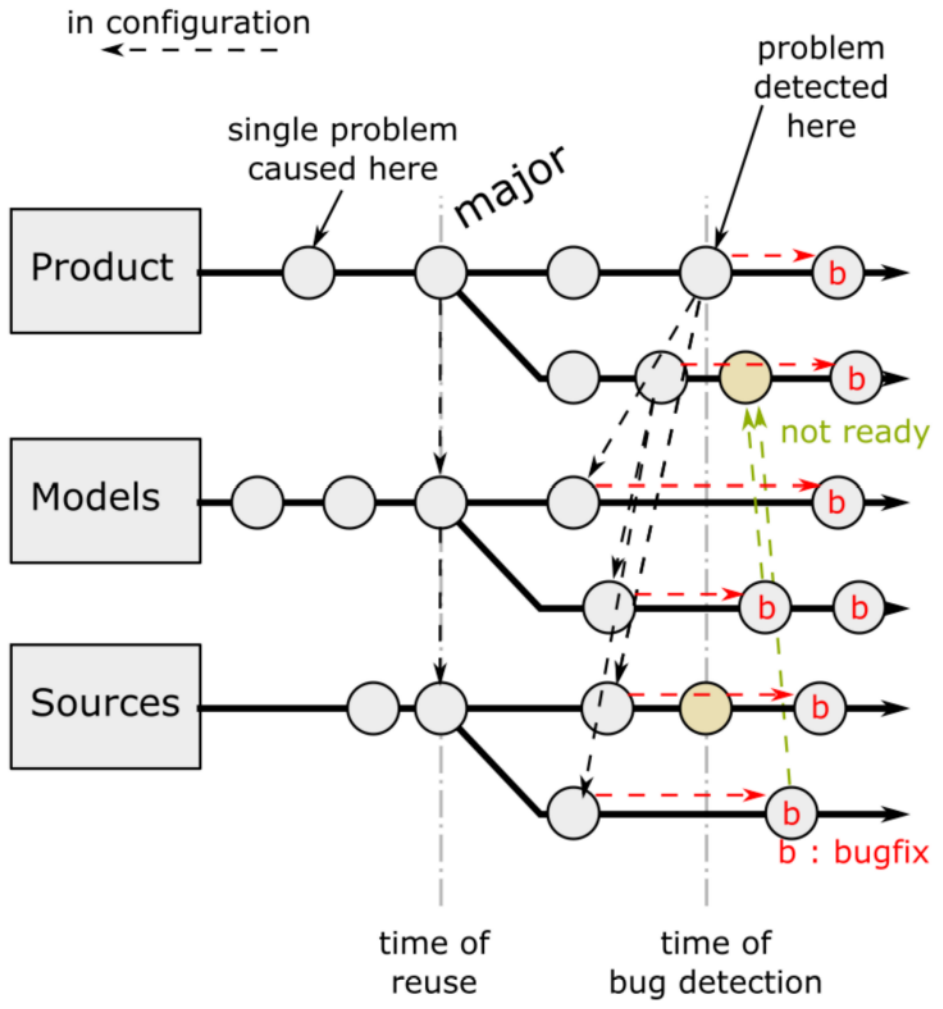

**Figure 4: The *bugfix nightmare* (brown baselines are skipped baselines which should have a bugfix but didn't manage to get it)**

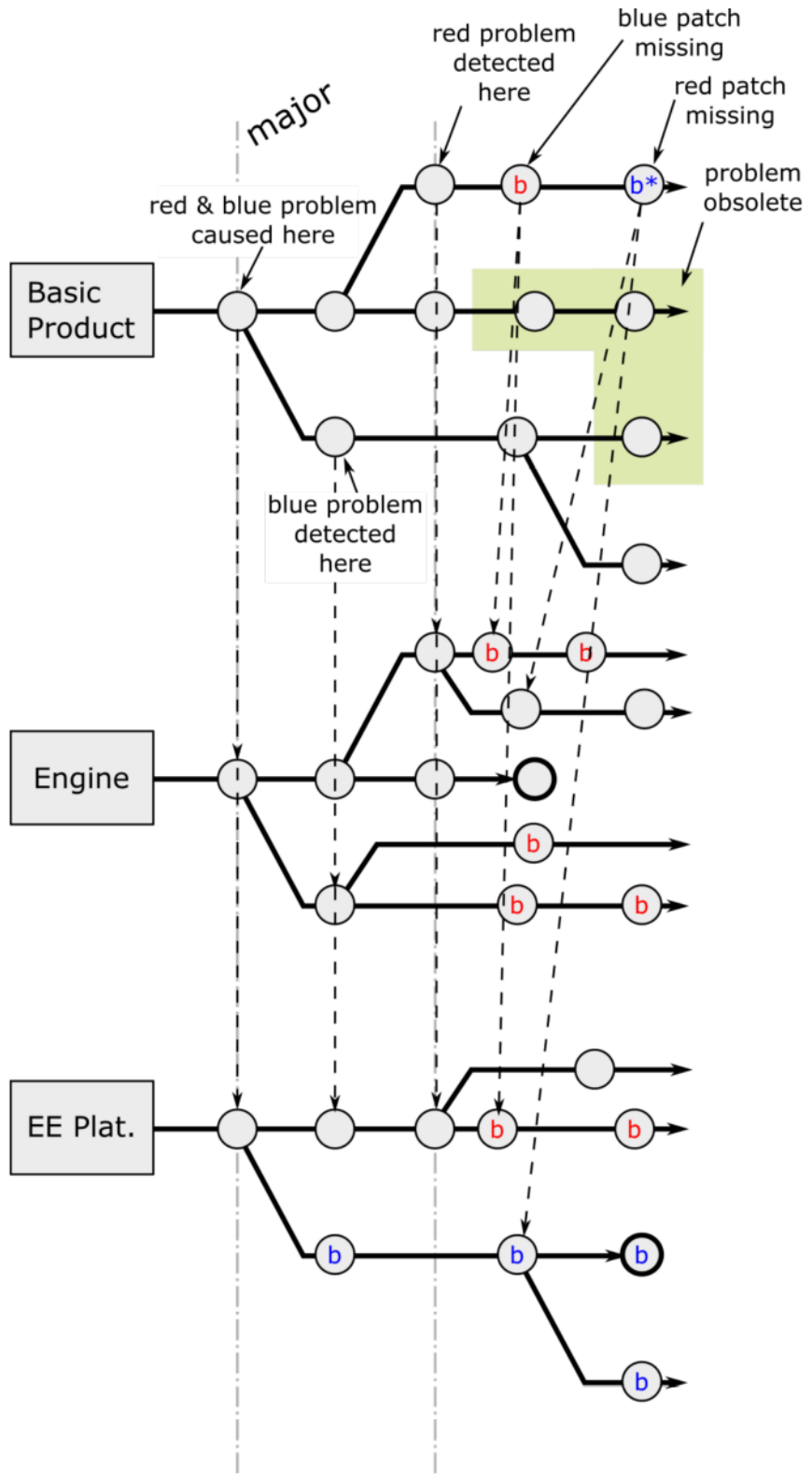

**Figure 5: The *bugfix hell*. Drawing all necessary arrows will render this picture incomprehensible.**

Once a sufficiently complex reuse scenario is reached, it is becoming more and more difficult to properly assign change requests or to evaluate impact from change requests up to the point where streams must be cut down or even a new family of product design is set up. Especially when changes related to certain bug-fixes (changes to changes) appear necessary, spotting them down is very tedious work.

What can be concluded so far? Unfortunately, baselines are hardly ever without flaw even with best effort to test, review and assess them. We hope it is intuitively clear that reuse realized by replicating design information is also replicating work related to fixing problems and that finding the right places to apply changes becomes less and less systematic over time.

As a consequence of this mundane fact we have questioned the correct choice of baselines as basic chain element of the stream. In fact we have the impression that it is rather motivated by straight forward technical concept feasibility and not so much by the needs of users. In order to fit the needs of users and not that of software developers we have replaced the baselines with bubbles. In the following chapter we would like to convince the reader that this specific choice has significant improvements for reuse even if it requires more sophisticated algorithms than that required for baselines.

That is why, in this paper the authors introduce the concept of *bubbles*, with the aim of providing a user-friendly concept to a collection of interacting, reuse-friendly mechanisms which can take advantage of existing services and enable the application of knowledge management techniques to boost reuse along the complete system development lifecycle.

In this light, bubbles emerge to play an important role in the overall design of the Technical Interoperability Concept (TIC, Figure 6) and to enable containerization of OSLC-based services. In the following, a very narrow description of the bubble concept is outlined, focusing on their application as a reuse mechanism, and will mainly concentrate on one question: *How do you manage reused design after it has been reused?*

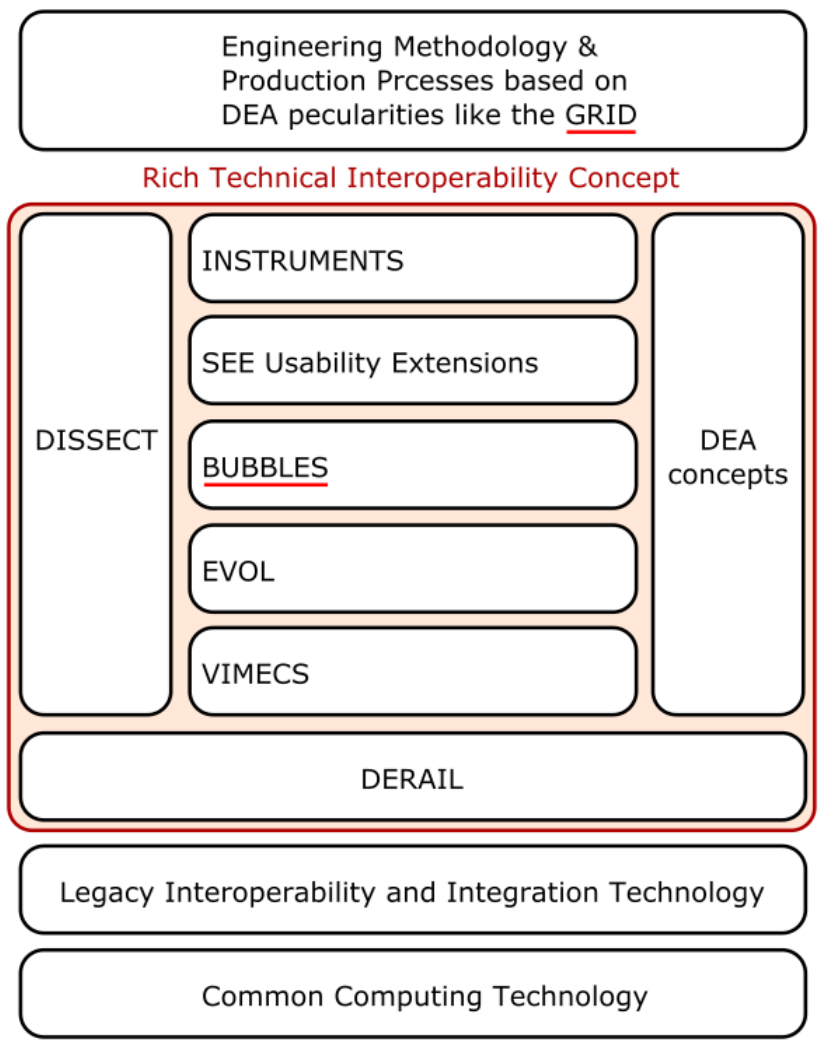

**Figure 6: Architecture of the Technical Interoperability Concept (TIC). Underlined elements highly contribute to advanced reuse practices.**

## Related work

In this section it is shown how bubbles relate to contemporary research trends on knowledge reuse and interoperability and how they uniquely combine best insights from them.

There seems to be two main lines of investigation. The first line is taking the approach to making data more universally available and to process it with general purpose processors. Taking into account that web technology is repetitively applied to systems and software reuse and that ever more solutions are explored on that ground for tackling the problems of storage, representation and retrieval, it seems that semantic approaches for the web can ease these tasks. In this light, the Semantic Web, as coined by Tim Berners-Lee in 2001 (Berners-Lee, Hendler, and Lassila 2001), has experienced a growing commitment during the last years from both academia and industrial areas with the objective of elevating the abstraction level of web information resources.

The Resource Description Framework (RDF), based on a graph model, and the Web Ontology Language (OWL), designed to formalize, model and share domain knowledge, are the two main ingredients to reuse information and data in a knowledge-based realm. Thus, data, information and knowledge can be easily represented, shared, exchanged [22] and linked to other knowledge bases through the use of Uniform Resource Identifiers (URIs), more specifically HTTP-URIs. As a practical view of the Semantic Web, the Linked Data initiative [12] emerges to create a large and distributed database on the Web by reusing existing and standard protocols. In order to reach this major objective the publication of information and data under a common data model (RDF) with a specific formal query language (SPARQL) provides the required building blocks to turn the Web of Documents into a real database or Web of Data.

OSLC is a clear representative of pursuing reuse via universal data formats and relatively standardized software functionality, as could be provided by the OSLC4J framework. Taking advantage of the Linked Data principles and Web standards and protocols, the OSLC effort emerges to create a family of web-based specifications for products, services and tools that support all the phases of the systems development lifecycle

Similar to OSLC, Agosense Symphony[3] offers an integration platform for application and product lifecycle management, covering all stages and processes in a development lifecycle. It represents a service-based solution with a huge implantation in the industry due to the possibility of connecting existing tools. WSO2[4] is another middleware platform for service-oriented computing based on standards for business process modeling and management. However, it does not offer standard input/output interfaces based on lightweight data models and software architectures such as RDF and REST. Other industry platforms such as PTC Integrity[5], Siemens Team Center[6], IBM Jazz Platform[7] or HP PLM[8] are now offering OSLC interfaces for different types of artifacts.

The other main line of research seems to focus on improved functional deployment which can then operate on specialized data. In recent times, we have seen the deployment of service oriented computing (Krafzig, Banke, and Slama 2005) as a new environment to enable the reuse of software in organizations. In general, a service oriented architecture comprises an infrastructure (e.g. Enterprise Service Bus) in which services (e.g. software as web services) are deployed under a certain set of policies. A composite application is then implemented by means of a coordinated collection of invocations (e.g. Business Process Execution Language). In this context, Enterprise Integration Patterns (EAI) (Hohpe and Woolf 2004) have played a key role to ease the collaboration among services. Furthermore, existing W3C recommendations such as the Web Services Description Language (WSDL) or the Simple Object Access Protocol (SOAP) have improved interoperability through a clear definition of the input/output interface of a service and communication protocol.

In the field of service-oriented computing, Cloud Computing systems and Service Oriented Architectures (SOA) have reached a level of complexity that implies the necessity of new methods and algorithms to automatically deal with the vast amount of data, variables, parameters, etc. that appear in this new realm in order to get an advanced management system (Jose María Alvarez-Rodríguez, Kourtesis, and Paraskakis 2014). The main problem lies in the complexity of designing models that allow an adequate management of a distributed architecture making decisions about resource provisioning, getting feedback for the final users, data and configuration management, etc. with the objective of avoiding existing "brute-force" solutions and overprovisioning. Furthermore, proper management of a

3 http://www.agosense.com/english/products/agosensesymphony/agosensesymphony

4 http://wso2.com/

5 http://www.ptc.com/application-lifecycle-management/integrity

6 http://www.plm.automation.siemens.com/en_us/products/teamcenter/

7 https://jazz.net/

8 http://www8.hp.com/us/en/business-services/it-services.html?compURI=1830395

service-oriented systems taking into account QoS (Quality of Service) is still an open issue and some relevant features such as saving costs, keeping high-performance, allocation of resources on-demand and offering a user-friendly experience to both IT managers and final users are still missing. In this sense, Autonomic Computing support for the next generation of cloud systems needs to be: 1) Self-x management, 2) agile, flexible and reliable, 4) deployable over a multiple cloud platforms, 5) handle complexity, 6) enable collaboration and coordination and 7) cost-effective and greener (energy-efficient). Under this context, semantic technologies have also emerged as an option to design and develop intelligent software components and to enable machines to automatically process and enrich data from different sources. Furthermore the sudden rise of containers (Bernhofen, El-Sahli, and Kneller 2013) or micro services leaded by solutions such as Docker[9] or Kubernetes[10] and other existing platforms like IBM Bluemix or Microsoft Azure has implemented a solution to overcome existing issues in cloud environments such as overprovisioning through the application of the well-known concept of containers in the Linux-based operating systems. Other solutions based on micro services such as Vagrant[11] enable users to share and create a flexible and adaptable development environment in a distributed context.

There have been attempts to explore mixes of the two lines of problem solving. In order to improve the capabilities of web services, semantics was applied to ease some tasks such as discovery, selection, composition, orchestration, grounding and automatic invocation of web services. The Web Services Modeling Ontology (WSMO) (Roman et al. 2005) represented the main effort to define and to implement semantic web services using formal ontologies. OWL-S (Semantic Markup for Web Services), SA-WSDL (Semantic Annotations for WSDL) or WSDL-S (Web Service Semantics) were other approaches to annotate web services, by merging ontologies and standardizing data models in the web services realm.

However, these semantics-based efforts did not reach the expected outcome of automatically enabling enterprise services collaboration. Formal ontologies were used to model data and logical restrictions that were validated by formal reasoning methods implemented in semantic web reasoners. Although this approach was theoretically very promising, since it included consistency checking or type inference, the reality proved that the supreme effort to create formal ontologies in different domains, to make them interoperable at a semantic level, and to provide functionalities such as data validation, was not efficient. More specifically, it was demonstrated (Rodríguez et al. 2012) that, in most of cases, data validation, data lifting and data lowering processes were enough to provide an interoperable environment. This insight is an important lesson taken for the design of bubbles. They provide clear context to data transport activities but also provide a relaxed system of data organization which prevents need of setting up infeasible meta-models of the data.

In the specific case of software engineering and reuse, the application of semantics-based technologies has also been focused in the creation of OWL ontologies (Castañeda et al. 2010) to support requirements elicitation, to model development processes (Kossmann et al. 2008) or to apply the Model Driven Architecture approach (Gaševic et al. 2006), to name just a few. These works leverage ontologies to formally design a meta-model and to meet the requirements of knowledge-based development processes. In contrast to these approaches, the

9 https://www.docker.com/

10 http://kubernetes.io/

11 https://www.vagrantup.com/

bubble absorbs practical orders provided by tools and the users are left to transform them into a more unified ontology later, when they see fit.

In conclusion, it is clear that systems reuse is an active research area that evolves according to the current trends in development lifecycles including knowledge management, service-oriented computing and micro services. It may have the potential of leveraging new technologies such as the web environment, semantics and Linked Data. However, data exchange does not necessarily imply knowledge management. From service providers to data items, a knowledge strategy is also required to really represent, store and search system artifacts, metadata and contents and enable automatic configuration of development environments.

In order to move forward new concepts are required. The bubble concept tries to adhere to contemporary technological developments, to learn from recent scientific insights and to strive for current industrial requirements. Its genericity and functionality has the promise to support various critical activities in development.

### *Motivation and Rationale*

The previous section has introduced the problem of providing flexible and reusable system development environments based on OSLC and knowledge reuse techniques. Following, a set of motivating questions are raised to evaluate existing technologies:

*$Q_1$: How do we clone a project with sources, models, requirements and tests in a convenient manner?*

This question aims at the missing containers as an interoperability concept. Without some kind of containerization it is very difficult to explicitly and precisely formulate any operation encompassing design information and hence its reuse.

*$Q_2$: How to reuse parts of designs without exaggerated duplication of their technical representation?*

*$Q_3$: How to make sure that no bug report or change request gets lost in a complex reuse scenario?*

The second and the third question aim at the avoidance of the negative downsides of duplicated data in reuse scenarios. The first problem is the huge amount of data produced. The second problem is the management of complex reuse flows.

*$Q_4$: How to manage reuse in scenarios of distributed storage?*

This question reminds of the fact that certain artifacts can only exist in certain environments. Irrespectively whether reuse means a copy or move, the artifact will never leave its natural habitat. From a technical perspective, copy or move must be implemented as separate concepts to “transport” and to enable reuse across heterogeneous, distributed environments.

*$Q_5$: How to realize reuse based on structural overlap?*

Finally, this last question is related to the idea that reused entities are not “convex hull items” but possibly open structures with various glue points which must be superposed. Reuse schemes of that kind occur when good structures must be reused and items play the role placeholders.

Currently, these questions can be quite reliably answered for different ubiquitous technologies that are summarized in Table 1 including [+1] if the challenge behind a question is easy to solve, [(+1)] if the question can be solved at a high cost (non-standard solution) or [0] if the question cannot be feasibly solved:

| Question | Files and Folders | SQL Databases | RESTful API (e.g. OSLC) |
|---|---|---|---|
| $Q_1$ | [+1] Easy if duplication. Otherwise limited | [0] Very difficult if at all possible | [0] No structural concepts exhibited |
| $Q_2$ | [+1] Use of links (feasibility depends on platform) | [+1] Very good. Databases can implement almost any reuse model (normalization). However, normalization is a developer activity. | [0] Theoretically possible if a "mapping" technique is used which is predominantly missing. |
| $Q_3$ | [0] Normally not possible. Files and folders are passive. It is possible to embed information in files and folders but formats must also support this. | [+1] Very good. Databases are quite good at maintaining active relationships. | [0] RESTful services are passive. Possible to write extra tools to emulate such features. Necessary features are not property of interfaces. |
| $Q_4$ | [(+1)] Remote folders can be mounted into larger structures (exotic features).<br>Not easy to manage. | [0] Very difficult but there is some development going on[12] (Tari 1992) (Brodie 1993, 1) (Oguz et al. 2015). | [+1] Very good. This is what the technology is designed for. |
| $Q_5$ | [(+1)] Possible to some extent with unifying file systems (exotic features).<br>Not easy to manage. | [0] SQL Databases don't do it. Some NoSQL Databases have such features [(+1)]. | [+1] The RESTful interface is not suited to this. However application of RDF can help to overcome deficiencies but OSLC does not fully exploit RDF capabilities. |
| **Reuse support estimation :** | 2/5 if we count the exotic features (4/5) | 2/5 (if we include some NoSQL Databases: 3/5) | 2/5 (if we count 1st and 3rd question as half completed, then 3/5) |

**Table 1: Preliminary evaluation of existing technologies to enable effective reuse.**

After this preliminary evaluation, it seems clear that none of the presented technologies were primarily designed to support reuse of design. Files and databases were designed as efficient, fast mass storage systems. Web-based interfaces, specifically the widely promoted RESTful APIs, are still targeting "transport activities" as should be expected from a network affine technology. Therefore, it is possible to conclude that it is necessary to define and implement a new hybrid technology that can serve the purpose of transport, conversion and

12 We do not mean in this context distributed databases which are based on sharding / horizontal splits. Those aim at scaling with distributed infrastructure. The task at hand is here that there is informational entanglement between objects or model segments of the overall ER-model which are run on different databases. In fact we even mean lazy modeling of overall ER-model and its exploitation.

storage while at the same time paying major concern to reuse. Here, a template for such technology is proposed under the term Technical Interoperability Concept (TIC). One of its outstanding features is the bubble which will be later evaluated following the same five questions.

## Bubbles: a data management approach to create an advanced industrial interoperability layer

Among the main goals of industrial interoperability projects like CRYSTAL, is to help the engineer to fully concentrate on design and to reduce his concerns for underlying data management and processing. From experience, this works the better the weaker the alignment is between the structures he can see and the technical structures used to run his tools. In some sense a technical Convey's law is holding true: You design the product structures only to approximate the structures favored by the tools. If you want to do something really innovative you must be able to move your design like if it did know nothing of files, folders or web services. A designer must be able to get the view of his design which is most likely to show him the problem and requires the least amount of edit in order to fix the problem. Developers who need not be concerned for technical entities anymore are only concerned for the logical structure of their work furthermore (see Figure 7). As a consequence they must be able to rely on basic operations on their logical structures irrespectively which tool feels responsible for doing the actual technical work.

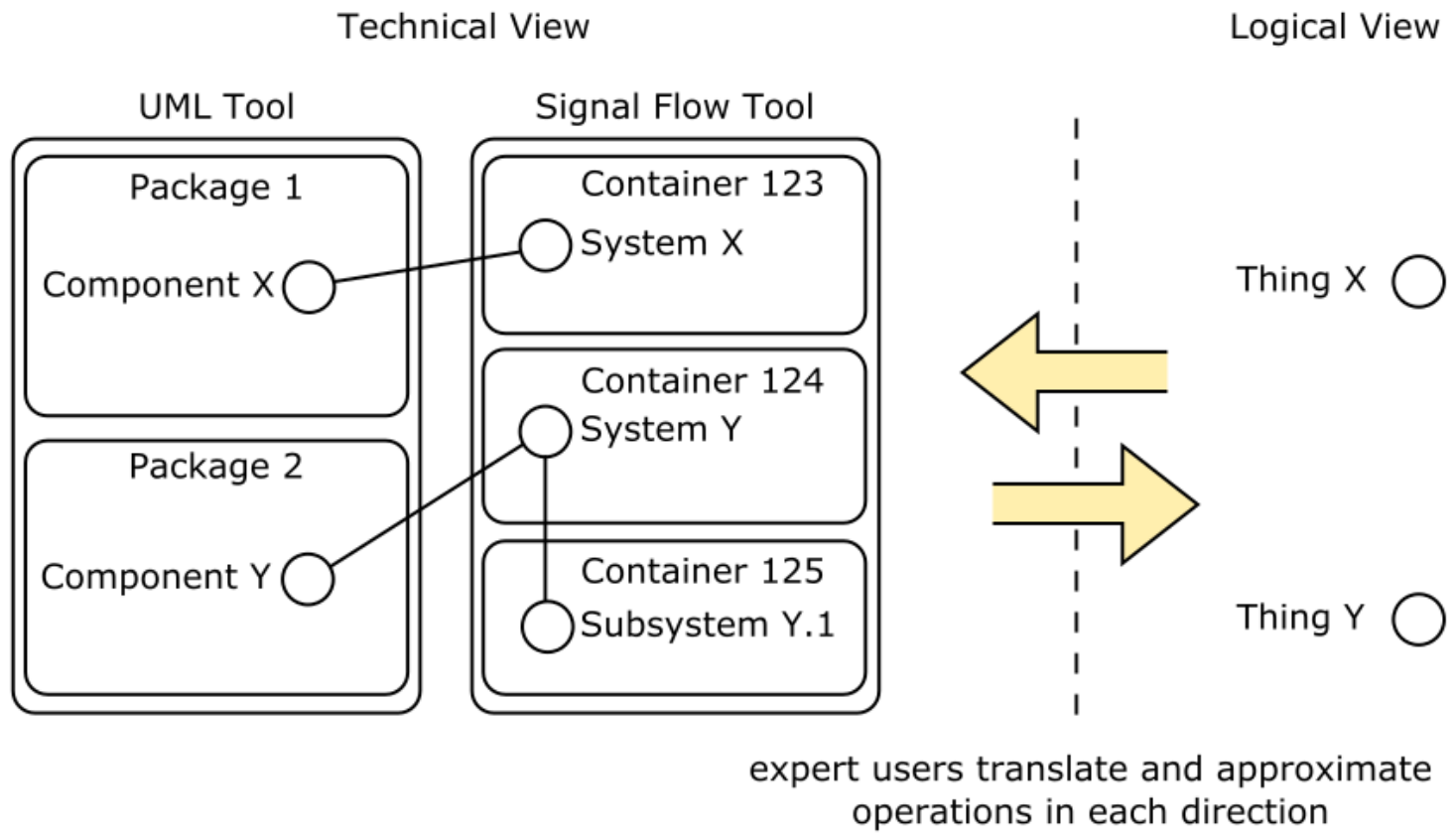

**Figure 7: Mental and technical structures can look vastly different. Performing coherent operations on the logical object requires a set of consistent operations in the technical view.**

But what to refer to in technical sense if anyone of the tools can just provide a technical artifact with his own set of rules, restrictions and possible operations? Actually, it can be concluded that the interoperability layer must manage his own descriptions of things and to relate it to any technical entity when demanded.

For this reason, a new concept unique to the interoperability layer is here introduced: *the Bubble*. The bubble is a logical parenthesis around other logical structures. Since Bubbles have been developed in order to support tools interoperability, it is a concept "on the line".

They act similarly to containers but there are also significant differences to traditional containers. One such difference is that it is highly structured while more common containers are just sets and it are only the relationships between those sets that yield structure. The Bubble maintains the appearance that it holds the data and storage that the tools are using. Despite this appearance bubbles are actually never the owners of the data or storage they are providing even if technical implementations of Bubble Governors or Bubble Proxies will cache some data for performance reasons. Bubbles can also contain portable functionality which is inheritable. This is because bubbles assume that they hold *objects* and not *data*. However, it is out of scope of this paper to treat this advanced feature.

With Bubbles a concept is introduced where the notion of place is the “virtual toolchain” – a highly abstract place that is not administered by the user at all. To his satisfaction she needs not to be concerned with the storage or transport details. In the TIC, the toolchain authorities and the IT authorities make sure to move data to new places. This is quite different from OSLC where the tool user has to configure her tool to work with certain services and where he is exposed with the actual physical distribution of data via the URIs.

According to the TIC, users are only concerned with creating Bubbles, deriving Bubbles and embedding Bubbles into each other – their operations remain in the purely abstract design space. Bubbles allow to segment this design space and are a fundamental contributor to reuse and distributed access control to data. For the user a bubble is a special reference to a design fragment, or more generically speaking to a knowledge fragment, which is very much behaving like a container and a real object: It can be used in different configurations with other things, It can be moved / ported, It can be created or destroyed, It can be frozen / snapshotted, It can be reproduced / cloned, It can be examined, It can be inserted into other objects.

These operations can be easily translated into major knowledge reuse activities, as all knowledge packets should be created, defined, indexed, found, altered, assembled and applied while respecting certain constraints all the time.

It is the job of the interoperability layer to resolve the abstract view provided by bubbles to technical terms at given time and location (workstation, network section, current server roles, etc.). In this context we point out the role of two further TIC concepts, the Extensible Virtual Object Layer (EVOL) and its interactions with Project Space Management (PSM), which are not covered here.

If implemented properly, the APIs to manipulating bubbles at this level are public, i.e. every tool is viable to initiate execution of these operations. Of course, there should be a default component allowing to inspect and manipulate bubbles even in the presence of tools which are not aware of any interoperability layers. Such default components should be provided as open-source free-of-charge software. Protection of distributed bubble operations is based on a (unfortunately) complex set of asymmetric encryption activities.

### *How bubbles help to reuse design*

Bubbles are unlike baselines. Bubbles are live containers for modeling design. The term *design* refers to a very wide concept. The content of the containers can be anything. In fact, Bubbles were designed to also manage physical[13] entities and to control their production. They are not primarily documenting historical relationships (like streams of baselines do) but functional / structural relationships, albeit users are free to abuse bubbles in order to model historical development. In fact, we would assume that users would first start using them like baselines and only recognize later that this is unnecessary. Bubbles do trigger snapshots for all involved elements in it but this version control process is perpendicular and possibly of very much reduced relevance once the users became acquainted with bubbles. Today, effective work demands high levels of competence in using version control systems. Users must translate between what they actually want and what the version control system is capable of doing. Advanced version control systems are not very well suitable to normal

13 Treating this interesting topic must be deferred to some future work.

users. No matter whether teams use centralized (e.g. Subversion) or distributed (e.g. Fossil) version control systems, none of them is doing what the bubble is doing for developers or users. We say that upfront because the following pictures will trigger associations to version control systems which will lead readers astray into false conclusions. Bubbles were designed to be usable by general users and they are not version control systems.

The most basic support for reuse offered by a bubble are the clone and derive operations. Figure 8 depicts two bubbles - the original bubble and a derived one. The derived bubble contains exactly the same information as the original but this was not obtained by copying any data. This has two advantages: The first advantage is that this operation is very fast. The second advantage is that the original and the derived design can be updated with a single edit. Only when the engineer decides to alter derived design it starts to create new technical resources which are uniquely becoming part of it. Editing a bubble implements the copy-on-write mechanism as it is known from filesystems.

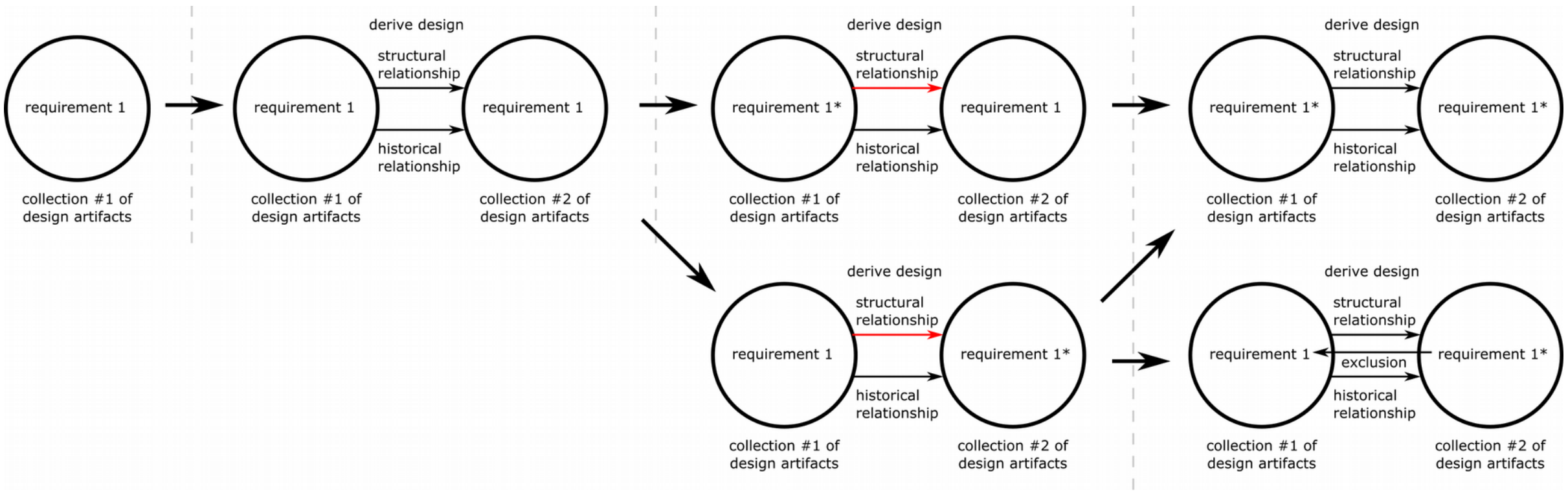

**Figure 8: Bubbles actively keep reused designs up to date. Technical representations are only duplicated when they are really different.**

This process is very similar to what PLM systems do with configurations. However, there are notable differences. Bubbles always feel like a singular cohesive item even if they contain nested bubbles. Unlike “configurations” bubbles indeed receive structural updates along the lines of inheritance which “configurations” do not. Data structures implementing “configurations” as we know them suffer from same problems of clone-and-own that they solve for their contents. On the contrary, bubbles inherit structural information the same way they inherit other content. This way the user can extend a “grandfather configuration” and have all descendants updated by the interoperability layer. To date, such configuration management functionality could be implemented in a tool but it would remain an isolated tool feature. Because bubbles are an interoperability concept even tools without any configuration management functionality can profit from it as long as they access data via the interoperability layer. More sophisticated tools which are providing configuration management models can derive new configuration data from the structural descriptions found in the bubble. For ALM/PLM systems the presence of bubbles implies that configuration descriptions will be influenced from external sources by interacting with the interoperability layer during development and not after.

Reuse with bubbles is not only something that is restricted to “large scale” items. There are large item reuse activities like reuse of a combustion engine in a new vehicle but also small reuses like replication of valve design in an array. Some authoring tools provide features to create reusable items while others will indeed rely on duplicates. Bubbles were designed to allow reuse from large to small scale but exploiting small scale reuse will rely on augmented tools which are aware of such features in the interoperability layer. However, we wanted to make the reuse hurdle for tools using such features very low. This resulted in two mechanisms: the cloning and the mirroring. These features serve not exclusively particular reuse activities but are generic operations of the bubble which have the nice side effect that tools can introduce templated design without having to introduce own complex algorithms in order to support it. It will be enough to augment certain copy and paste operations with interactions with the interoperability layer. In Figure 9 you can see a picture showing the clone operation which is creating two different relationships: a historical and a structural. The example shows, how the bubble represented by collection #2 has been extended with a refined requirement 1a. After cloning bubble #2 into bubble #3 a misalignment of historical and structural relationships takes place which is intended. After requirement 1a has been updated to 1b the technical reuse of bubble #2 has ended and the historical relationship remains purely for documenting purposes.

Why is bubble #3 maintaining reuse relationships to two bubbles? The goal of the TIC is to eliminate duplication of data as much as possible and this implies the reuse of bubble #2 as much as possible. The bubble represented by collection #2 is containing inherited, overwritten and original design elements. Without inheritance from bubble #2, bubble #3 would either contain duplicates or would not contain the refined requirement. Because there are two different relationships involved, it is possible to develop the third bubble either in co-step with the second or to slowly dissolve its entanglement with it without ever influencing the correct inheritance from the original source bubble (represented by collection #1). The correct inheritance of all knowledge management items and changes is guaranteed even after the historical relationship has become functionally meaningless.

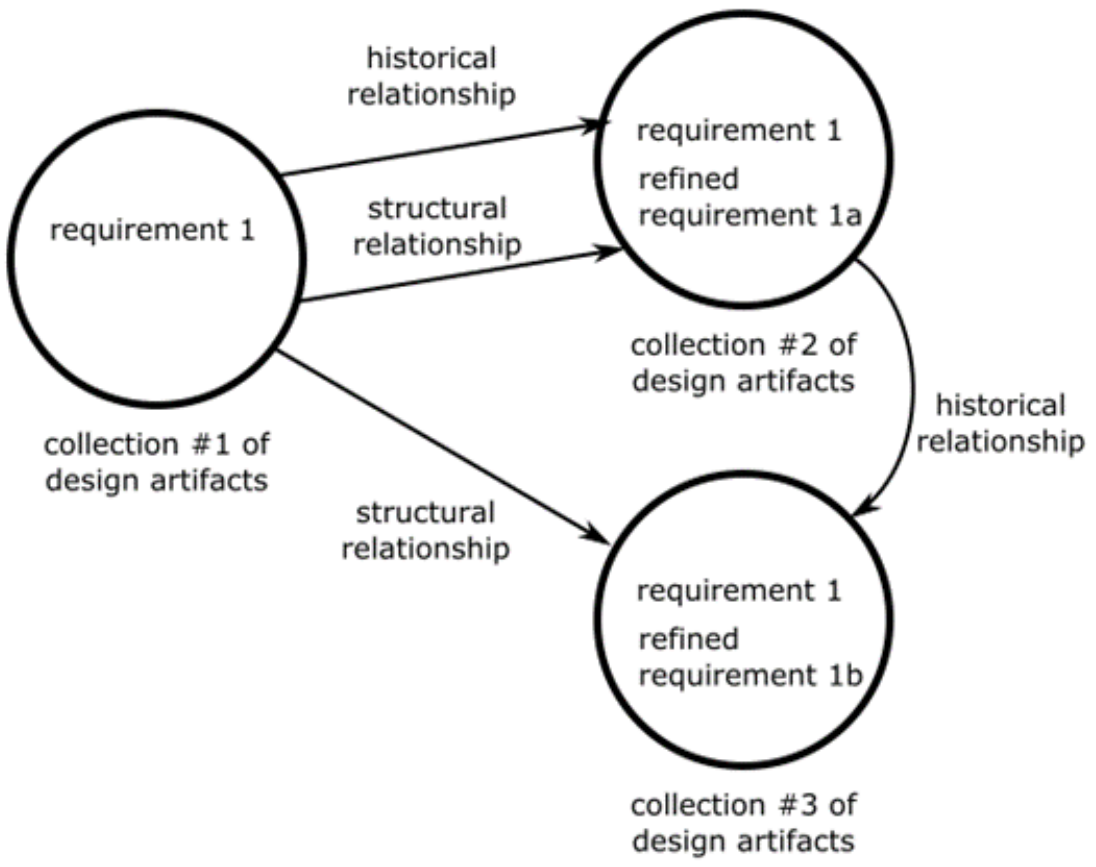

**Figure 9: Operations on Bubbles involving creation leave relationships between them which do not rely on a specific modelling capability of any toolchain party.**

Despite the technical synonymy between cloning and derivation, the two different operations are necessary. The difference in resulting link-up will positively influence design stress signaling paths between the bubbles when development continues. Also, unlike derived bubbles, cloned bubbles need not respect constraints management provided by the source bubble – it can define its constraints freely, only controlled by its structural ancestor.

### Case Study: Simplifying bug fixes using Bubbles

Since the stream of bubbles represents lines of reuse and because bubbles are live design containers it is now easy to imagine how problems will be solved in the most natural way possible (see Figure 10). The caretaker of the problem will inspect ancestorship of the bubbles until he finds the relevant design to be altered. On the way he will find all relevant design decisions in the respective knowledge bases so that he can decide how far he wants to go back in order to solve the problem. He can clearly see in advance which other projects will be affected and to summon a meeting in order to discuss the effect of planned change on them. In this process the work that is today considered difficult is a relatively trivial operation for the interoperability layer based on bubbles.

The applied bug fix can be simple or complex. A simple fix would be an update in a source file. Let's quickly assume that this source file is not actively edited anymore. By committing the file to the bubble a single technical resource is updated which is referenced by all subsequent bubbles in the stream. The bubble is sending a *design stress* signal to derived and cloned bubbles which can at minimum use this occasion to refresh bubble descriptors or to trigger bubble-specific event handlers. Such event handler could notify a bubble user that there has been an upstream design change and if he wants to accept it. In the regular case all bug fixes and changes follow the exact path of reuse as can be seen in Figure 10. This is even true when this change is consisting of updates made of heterogeneous data, web resources or files, or even a structural update of how the bubble is organized.

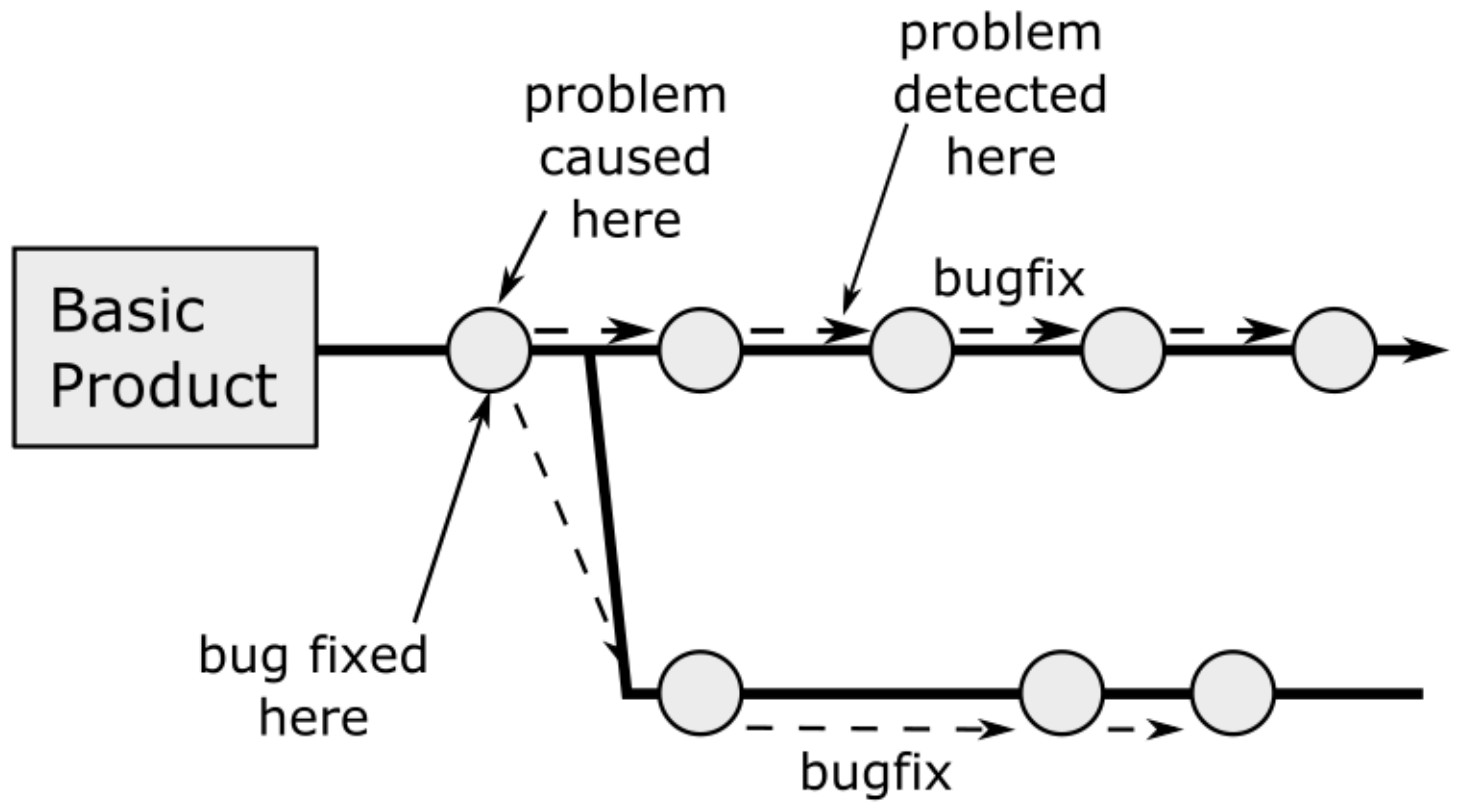

**Figure 10: The faulty design will be fixed first. Inheriting projects receive bugfix in a totally straight forward manner along all documented clone and derive relationships. This is also true for nested bubbles.**

If the user declines then the propagation of stress signal to further bubbles down the stream is interrupted and a reverse signal is sent. This signal must create an inserted bubble which is based on data from the versions before the change. That's the reason why the bubble will only signal after commit. This situation is exploiting the special capabilities of bubbles to insert or retract intermediate designs and to emit and consume signals among them (see Figure 11). In some ways this is similar to a rebase action as they are known in version control systems but the similarity is superficial. It will depend on the intelligence of bubble governors how it can prevent superfluous insertions. However, since bubbles only point to the very next bubble, insertions and removals are as easy as with a traditional pointer chains.

If all affected users accept the changes then all bubbles will point to exactly the same technical resource which is reused in all the bubbles. Fixing a resource which has been already edited will require the actual merge. At least the manual merge should be always possible. For this to work the activated bubble which has been signaled of suffering design stress must have the means to ask the user to resolve stress by performing a certain merge. If a bubble is activated by anyone then the stress signal is remembered. It will simply take effect the next time one of its legitimate users activates it.

The more interesting cases occur if the design patch is not to be accepted. For example, accepting a source file patch is probably easy. Accepting a reorganization of bubbles is probably more difficult. What would occur in such case? In order to demonstrate how the design space gets reorganized we have prepared an example sequence to be seen in Figure 12 to Figure 17. The rebasing process is taken along the stream to each derived bubble. The teams decide if they want to go on the new stream or stay on the old. In any case their current design gets cloned. In this process we see superfluous designs to be created that no team would like to take care of. If this is the case then those bubbles will be retracted and the content merged with follow up bubbles so that nothing gets lost. After all operations have been performed we see a new design hierarchy. This new hierarchy is subdivided in design stream with a red feature and one without. Because both streams still inherit from a large common body of design they will also receive stress signals regarding this body of design (or body of knowledge) irrespectively where the problem will be found in the hierarchy. The process of change application remains straight forward – it will occur along the now existing derivation relationships.

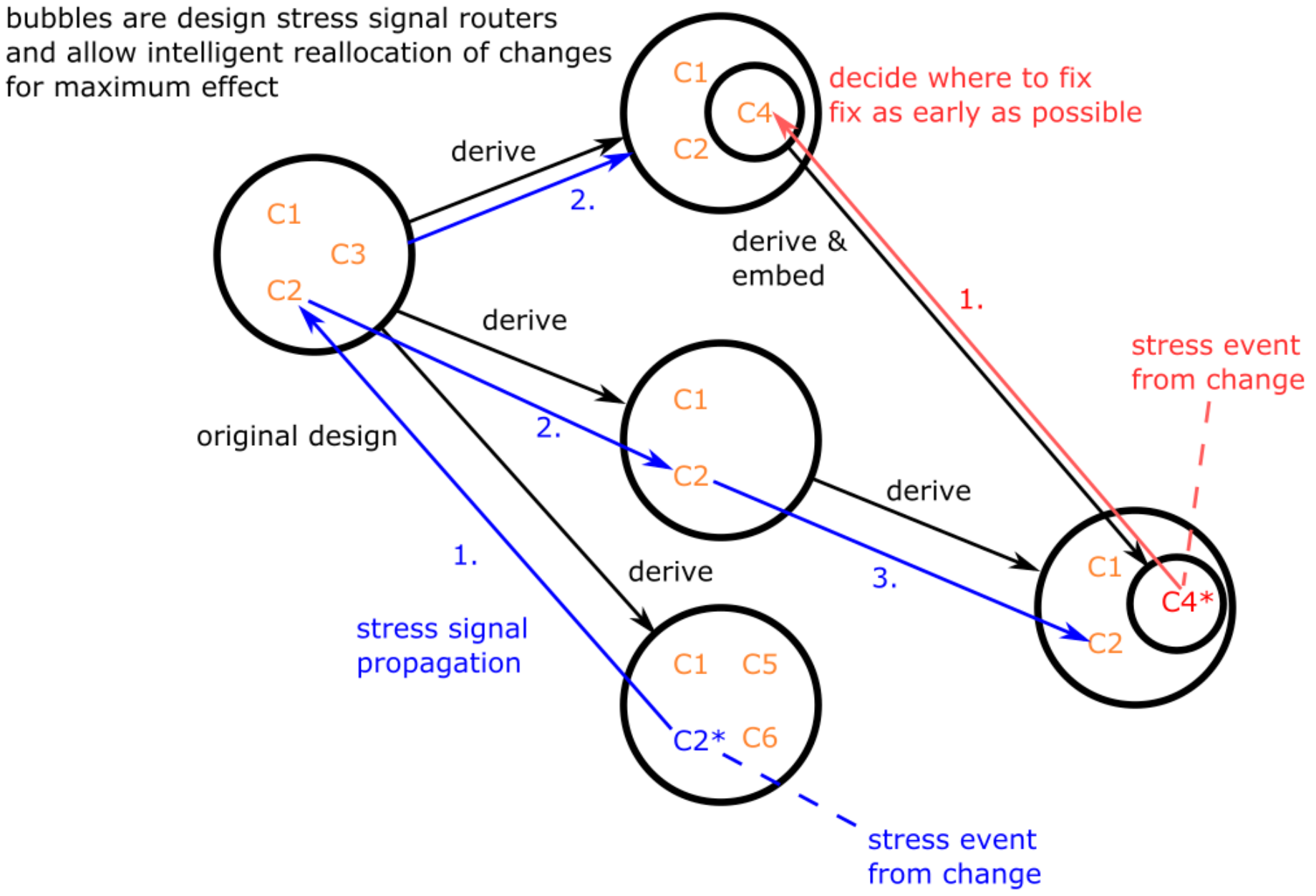

**Figure 11: Bubbles are design stress signal routers.**

We can imagine that it is important for developers to know how much stress is on a certain bubble stream. This would help him to anticipate future changes. In order provide this kind of overview bubbles emit uninterruptible stress signals upstream and downstream. If a developer activates a bubble showing high stress levels then he might decide to refrain from editing it because he can assume that system architects are performing grave transformations of design which will later affect him. If he wants to know more which changes are performed then he can access the descriptive parts of the bubble which have been inherited along derivation or embedding lines. This way he can know who is dealing with the overall design. What he could do is to rebase the stream to an inserted bubble like we have seen before. In that case he communicates clearly that he does not want to participate from updated architecture.

This operation cannot be easily reverted without causing a lot of merging operations on the bubble stream. This might look like a "problem" but is indeed intended. The goal of the bubbles is to detect irrelevant or even harmful work and to prevent it. The goal is not to keep teams occupied with additional work resulting from harmful work. Cost of reuse can only be reduced if teams get clear signals where to do beneficial work that does not produce high follow up efforts resulting from maintenance and merging. However, nothing prevents developers to branch into private bubble streams but their work will be clearly isolated and eventually not mandated.

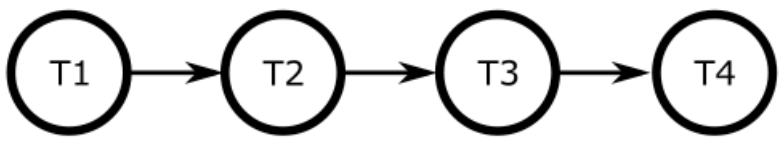

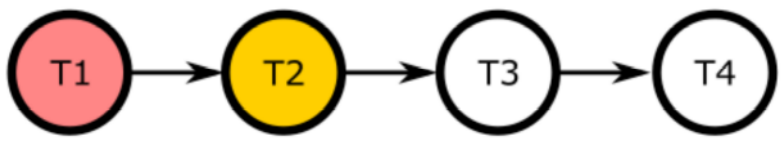

**Figure 12: Design stream (Bubble stream) with no stress. T1...T4 are teams.**

**Figure 13: T1 introduces a change that would affect derived designs. Red is feature. Yellow is stressed bubble.**

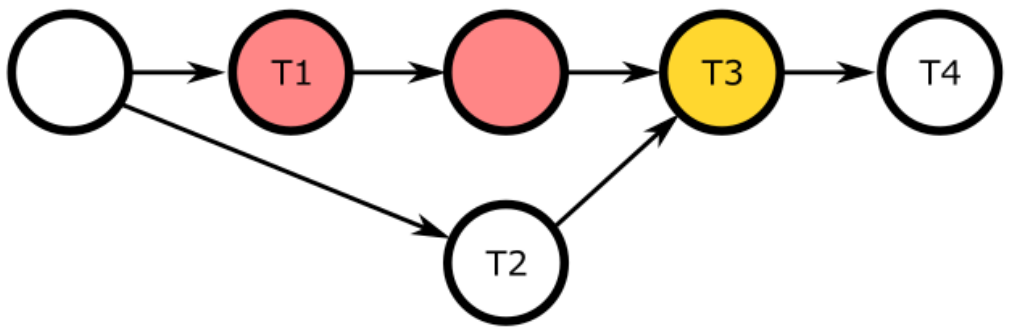

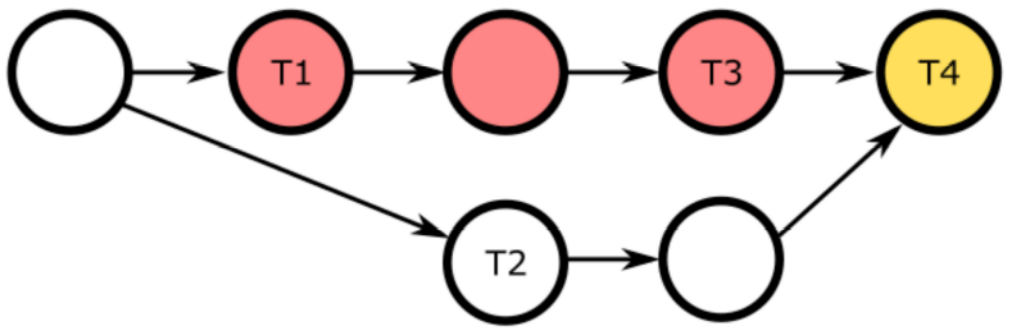

**Figure 14: Team T2 decides not to follow design change and puts T3 at stress.**

**Figure 15: Team 3 decides to accept design change and stay on red stream.**

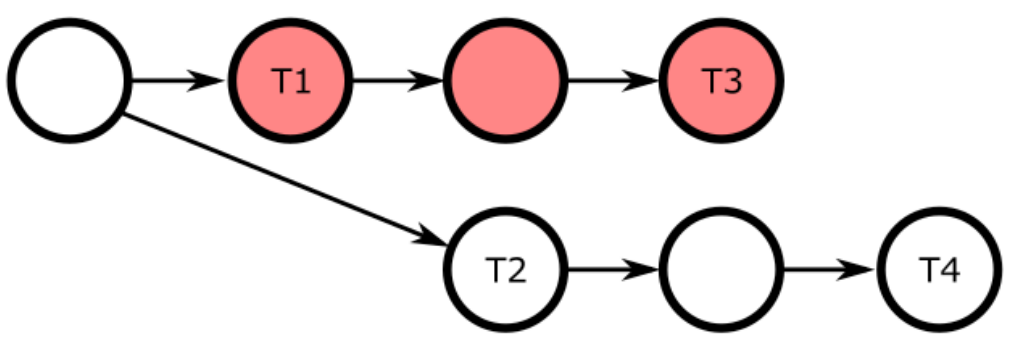

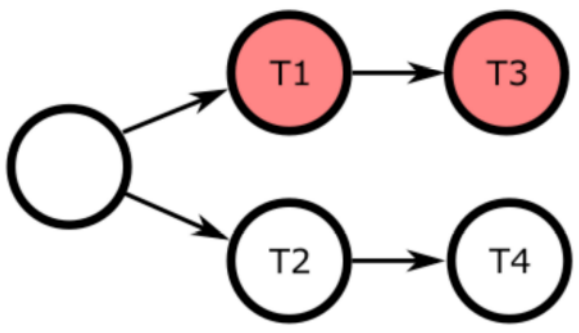

**Figure 16: Team T4 decides not to accept design change. This results in two design lines: red & white.**

**Figure 17: After irrelevant intermediate designs have been retracted the overall design architecture of the product lines has changed.**

Bubbles provide an embed operation if parts from the private branches shall be reused but additional mechanics of the bubbles not described in this paper will detect violations of assumptions during such merges. It can very well be that the project leader will then not allow reuse of such component, especially in projects developing critical systems (see Figure 18).

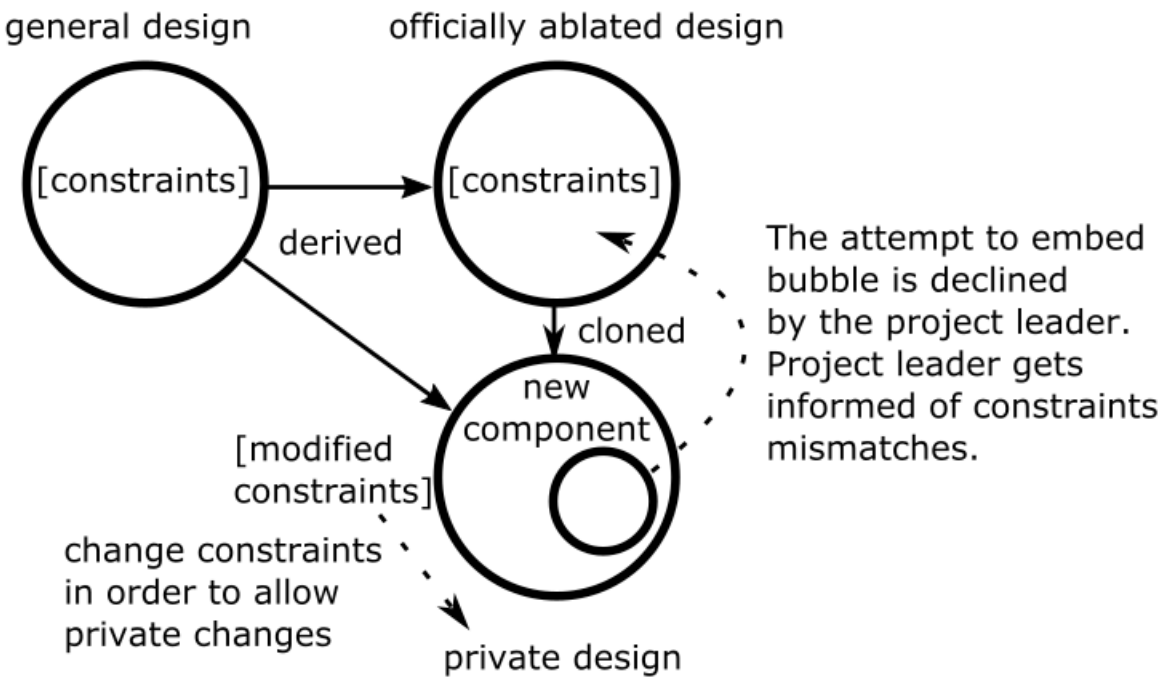

**Figure 18: Prevention of illegal reuse with bubbles. Bubbles have explicit notion of design constraints.**

## Evaluation and Discussion

This work took the opportunity to motivate bubbles and some of the intention behind it and it gave a first look into mechanisms associated with a bubble. The main contribution is to show bubbles as reusable items in broadest sense, see Table 2. In several places, it has been remarked that it is probably difficult for the first time reader to understand how the bubble is different from the concepts he is familiar with. Probably, one good way to learn about a new concept is to discriminate it clearly from other concepts. We will quickly compare the bubble paradigm with other paradigms in order to highlight its unique properties.

- **A bubble is not a version like it would be implemented in a version control system.** You can insert bubbles between bubbles or retract them. Revisions cannot be altered but bubbles can. Bubbles allow to create, manage and refactor your design architecture. In many ways bubbles are more similar to classes or objects in object oriented programming languages than they are similar to versions or revisions. What bubbles do in terms of version management is to initiate distributed snapshots and to provide a contextual view to versioned items in version control systems. Bubbles use version control systems more like backup facilities.
- **A bubble is not a variant**. Despite that major streams of bubbles can be understood as variants this is not applicable to all situations where you would introduce new bubbles. Bubbles help to organize variant spaces and do include in their conception mechanisms to do this efficiently beyond current state-of-the-art. In fact, with bubbles you cannot even have no variant management in your toolchain. It is possible to abuse bubbles to the point where each bubble is synonymous with a variant but this use falls short of bubble capabilities. We are convinced that nobody in engineering, production or support is directly caring of variants if he has a more precise concept to identify information that he is needing.
- **A bubble is not a configuration - not a configuration like is understood by PLM systems.** It is not used to collect and assemble components in specific versions. This activity is just the result of insufficient concepts for tooling and interoperability. It is true that bubbles aggregate other bubbles but they do so for the reason of reuse and not for the reason to overcome limited snapshotting capabilities of toolchain. In a static view a bubble can look like a configuration but the dynamics of bubbles are different.
- **A bubble is not a baseline.** Baselines are just special versions of versions or revisions. Baselines want to provide a stable data ground for a team to work on. In order to do so they are normally only created after some quality has been proven. Finding deficits in data to be baselined can quickly lead to a branch if work has to continue on the main line of development. After applying a set of corrections on a baseline, it can be found at the end of a special branch in a version control system. Downsides of the process is that fixes to the branch are not automatically forwarded towards the trunk. With bubbles the process is quite different. The slow-down party inserts a bubble in a bubble stream where it will fix quality which is immediately inherited. The existence of this "high quality" bubble becomes mostly irrelevant after the fact. If developers still feel like they need a "baseline" they can create a short bubble stream which is terminated. But what would be any good reason to use it ever after?

This brings us to the last question, whether bubbles are like branches in version control systems. This is probably the greatest similarity with conventional terms because you would branch a bubble stream and a version stream if you wanted to do isolated exploration or development of design. But this is where similarity probably ends. Continuous edits of a bubble will not create a stream of bubbles like it would happen with a version control system if a developer keeps editing his branch.

Differences also exist in how and under what circumstance merging takes place. The process of merging has different meaning for bubbles and for branches. Merging conventional branches reduces the number of branches. The merging operation results in a new common feature but only after the head revision. In bubbles, including a new feature will affect the whole appended stream of bubbles, no matter how wildly reused.

The number of actual merge activities is indefinite. In best cases there will be no merges at all between bubbles. In worst cases propagating features will cause several diminishing merges of design. In this process we have seen situations of a design stream rip off which leads to an improved overall design architecture. Since bubbles are a distributed concept they are equipped with a special merging function which becomes ever active when there is delay on the line during their synchronization. But since bubbles do not hold the data like version control systems the process looks different. Even if the exact details are not clear to the reader right now, we would like the reader to simply keep in mind: Neither bubbles nor bubble streams do behave like branches known from version control systems.

Finally, bubbles should be compared with existing technologies using the five questions schema introduced in Motivation and Rationale. Since bubbles have not yet been demonstrated as implementations, that comparison is purely conceptual and might appear unfair for existing technology. However, operations defined for the bubble have all implemented representatives in real systems and that is why the concept is not too fantastic. It is the mixture of features and the allocation of responsibilities to the interoperability layer that is new. Table 2 summarizes how well bubbles should be able to support reuse activities given the five questions. At current state of analysis bubbles should be an adequate concept for facing future challenges in reuse practice.

| **Question** | **TIC (Bubbles)** |
|---|---|
| Q1 | [+1] Easy in every sense. Cloning design and rearranging parts without constraints is the main feature. |
| Q2 | [+1] Bubbles implement lazy copy-on-write mechanics |
| Q3 | [+1] Bubbles are active signal routers which can maintain active relationships and information related to reuse path. |
| Q4 | [+1] Bubbles emulate a highly dimensional structured storage which expands upon need |
| Q5 | [+1] Bubbles use namespaces and structures which are auto-merged along embedding relationships. This provides a universal overlay mechanism for the whole toolchain. |
| Reuse support estimation: | 5/5 |

**Table 2: Preliminary evaluation of Bubbles following the questions presented in Table 1.**

## Conclusions and Future Work

In this paper the main motivations and basic notions of the Bubble have been presented. The Bubble is an important key concept in the overall design of the TIC. Bubbles are a new concept with confusing similarities to other known concepts but also with significant differences. Since there exists no product which implements TIC functionality (and hence bubbles) it is too early to demonstrate bubbles or evaluate their governing components' performance on some tasks. Frankly, the extended goal of this paper is to assemble interested parties from research and industry in order to provide first reference implementations of an Advanced Industrial Interoperability Layer (AIIL) based on TIC principles as have been first laid out during the CRYSTAL project. Our current approach would be to build it in an as OSLC-friendly way as possible in order to exploit current impulse in industry to standardize interoperability features. However, AIIL greatly exceeds goals set by OSLC or CRYSTAL IOS, by scope, effect and function.

Bubbles were designed with a broad range of applications and activities in industrial environments in mind. They are a corner stone to the idea of managing design from inception to support of products in field and for facing growing autonomy of products and production facilities. All concepts of the TIC were clearly geared towards effectively facing future challenges like Industrial Internet, Internet of Things, Internet of Tools or Economies based on an Autonomous Production Fabric. Various industry silos start to develop their next generation interoperability technology. Once these silos become too consolidated we see the risk that reaching innovative industry features will become inhibited by avoidable technology segmentation. Why not anticipate the inevitable and not design interoperability and integration technology which would survive not only the fourth but also the fifth industrial revolution? This was our challenge.

The resulting TIC relies on unfamiliar concepts which require rethinking of many engineering and production challenges by readers exposed to it for the first time. The downside to this fresh awkwardness is that it is impossible to cover the full spectrum of impact on data management and engineering methodologies in the scope of a single paper. In order to get started we have motivated TIC's bubbles with a very narrow view on them by focusing on how they help to reuse artifacts in product lines. Further work may include the full implementation of the bubbles-view on top of existing tools such as Git, distributed filesystems and micro-services and their complete theoretical definition.

### *Acknowledgements*

The research leading to these results has received funding from the ARTEMIS Joint Undertaking under grant agreement Nº 332830-CRYSTAL (CRitical sYSTem engineering AcceLeration project) and from specific national programs and/or funding authorities. This work has been supported by the Spanish Ministry of Industry and by the Germany Ministry of Education and Research (BMBF, 01IS13001E).

## References

Alvarez-Rodríguez, Jose María, Dimitrios Kourtesis, and Iraklis Paraskakis. 2014. "Semantic-Based QoS Management in Cloud Systems: Current Status and Future Challenges." *Future Generation Comp. Syst.* 32: 307–23.

Alvarez-Rodríguez, Jose Maria, Juan Llorens, Manuela Alejandres, and Jose Fuentes. 2015. "OSLC-KM: A Knowledge Management Specification for OSLC-Based Resources."

In *Proceedings of the 25th Annual INCOSE International Symposium (Accepted)*. Vol. 25.

Benavides, David, Sergio Segura, and Antonio Ruiz-Cortés. 2010. "Automated Analysis of Feature Models 20 Years Later: A Literature Review." *Information Systems* 35 (6): 615–36. doi:10.1016/j.is.2010.01.001.

Berners-Lee, Tim, James Hendler, and Ora Lassila. 2001. "The Semantic Web." *Scientific American* 284 (5): 34–43.

Bernhofen, Daniel M, Zouheir El-Sahli, and Richard Kneller. 2013. "Estimating the Effects of the Container Revolution on World Trade."

Biggerstaff, T. J., and C. Richter. 1989. "Software Reusability: Vol. 1, Concepts and Models." In , edited by Ted J. Biggerstaff and Alan J. Perlis, 1–17. New York, NY, USA: ACM. http://doi.acm.org/10.1145/73103.73104.

Boehm, Barry W. 1981. *Software Engineering Economics*. 1st ed. Upper Saddle River, NJ, USA: Prentice Hall PTR.

Brodie, Michael L. 1993. "The Promise of Distributed Computing and the Challenges of Legacy Information systems1." In *Interoperable Database Systems (Ds-5)*, edited by DAVID K. HSIAOERICH J. NEUHOLDRON SACKS-DAVIS, 1–31. IFIP Transactions A: Computer Science and Technology. Amsterdam: North-Holland. http://www.sciencedirect.com/science/article/pii/B9780444898791500075.

Castañeda, Verónica, Luciana Ballejos, Laura Caliusco, and Rosa Galli. 2010. "The Use of Ontologies in Requirements Engineering," Global Journal of Researches In Engineering, 10 (6). http://engineeringresearch.org/index.php/GJRE/article/view/76.

Davis, Randall, Howard Shrobe, and Peter Szolovits. 1993. "What Is a Knowledge Representation?" *AI Magazine* 14 (1): 17.

Eran, Gery, and Joanne L. Scouler. 2014. "Strategic Reuse and Product Line Engineering – Using IBM Rational Systems and Software Engineering Platform." *IBM Corporation*.

Gaševic, Dragan, Vladan Devedžic, Dragan Djuric, and SpringerLink (Online service). 2006. *Model Driven Architecture and Ontology Development*. Berlin, Heidelberg: Springer-Verlag Berlin Heidelberg. http://proxy.library.carleton.ca/login?url=http://dx.doi.org/10.1007/3-540-32182-9.

Groza, Tudor, Siegfried Handschuh, Tim Clark, S Buckingham Shum, and Anita de Waard. 2009. "A Short Survey of Discourse Representation Models."

Hohpe, Gregor, and Bobby Woolf. 2004. *Enterprise Integration Patterns: Designing, Building, and Deploying Messaging Solutions*. The Addison-Wesley Signature Series. Boston: Addison-Wesley.

Hull, Richard, and Roger King. 1987. "Semantic Database Modeling: Survey, Applications, and Research Issues." *ACM Computing Surveys (CSUR)* 19 (3): 201–60.

INCOSE. 2004. "Systems Engineering Vision 2020." Technical INCOSE-TP-2004-004-02. INCOSE. http://www.incose.org/ProductsPubs/pdf/SEVision2020_20071003_v2_03.pdf.

Jacobson, Ivar, Martin Griss, and Patrik Jonsson. 1997. *Software Reuse: Architecture, Process and Organization for Business Success*. New York, NY, USA: ACM Press/Addison-Wesley Publishing Co.

Jose María Alvarez-Rodríguez, Juan Llorens, Manuela Alejandres, and Jose Fuentes. 2014. "Towards a Semantic-Based Representation and Computation of Quantitative Indexes for Quality Management of Requirements." In *Proceedings of the 24th Annual INCOSE International Symposium (Accepted)*, 24:27–40. doi:10.1002/j.2334-5837.2014.tb03132.x.

Karlsson, Even-André, ed. 1995. *Software Reuse: A Holistic Approach*. New York, NY, USA: John Wiley & Sons, Inc.

Kim, Yongbeom, and Edward A. Stohr. 1998. "Software Reuse: Survey and Research Directions." *J. Manage. Inf. Syst.* 14 (4): 113–47.

Kossmann, Mario, Richard Wong, Mohammed Odeh, and Andrew Gillies. 2008. "Ontology-Driven Requirements Engineering: Building the OntoREM Meta Model." In , 1–6. IEEE. doi:10.1109/ICTTA.2008.4530315.

Krafzig, Dirk, Karl Banke, and Dirk Slama. 2005. *Enterprise SOA: Service-Oriented Architecture Best Practices*. Prentice Hall Professional.

Krueger, Charles W. 1992. "Software Reuse." *ACM Computing Surveys (CSUR)* 24 (2): 131–83.

Manikas, Konstantinos, and Klaus Marius Hansen. 2013. "Software Ecosystems – A Systematic Literature Review." *Journal of Systems and Software* 86 (5): 1294–1306. doi:10.1016/j.jss.2012.12.026.

Marko, Nadja, Grischa Liebel, Daniel Sauter, Aleksander Lodwich, Matthias Tichy, Andrea Leitner, and Jörgen Hansson. 2014. "Model-Based Engineering for Embedded Systems in Practice." 2014:01. Research Reports in Software Engineering and Management. Gothenburg: University of Gothenburg. https://gupea.ub.gu.se/handle/2077/37424.

Mcilroy, Doug. 1969. "Mass-Produced Software Components." In *Proceedings of Software Engineering Concepts and Techniques*, edited by J. M. Buxton, P. Naur, and B. Randell, 138–55. Garmisch, Germany: NATO Science Committee. http://homepages.cs.ncl.ac.uk/brian.randell/NATO/nato1968.PDF.

Mili, Hafedh. 2002. *Reuse Based Software Engineering: Techniques, Organization and Measurement*. New York: Wiley.

Mili, Hafedh, Fatma Mili, and Ali Mili. 1995. "Reusing Software: Issues and Research Directions." *Software Engineering, IEEE Transactions on* 21 (6): 528–62.

Nonaka, Ikujiro, and Hirotaka Takeuchi. 1995. *The Knowledge-Creating Company: How Japanese Companies Create the Dynamics of Innovation*. New York: Oxford University Press.

Oguz, Damla, Belgin Ergenc, Shaoyi Yin, Oguz Dikenelli, and Abdelkader Hameurlain. 2015. "Federated Query Processing on Linked Data: A Qualitative Survey and Open Challenges." *The Knowledge Engineering Review* 30 (05): 545–63. doi:10.1017/S0269888915000107.

Prieto-Díaz, Rubén. 1991. "Implementing Faceted Classification for Software Reuse." *Commun. ACM* 34 (5): 88–97. doi:10.1145/103167.103176.

Rodríguez, Miguel García, Jose María Alvarez-Rodríguez, Diego Berrueta Muñoz, Luis Polo Paredes, José Emilio Labra Gayo, and Patricia Ordóñez de Pablos. 2012. "Towards a Practical Solution for Data Grounding in a Semantic Web Services Environment." *J. UCS (JUCS)* 18 (11): 1576–97.

Roman, Dumitru, Uwe Keller, Holger Lausen, Jos de Bruijn, Rubén Lara, Michael Stollberg, Axel Polleres, Cristina Feier, Cristoph Bussler, and Dieter Fensel. 2005. "Web Service Modeling Ontology." *Applied Ontology* 1 (1): 77–106.

Smolárová, Mária, and Pavol Návrat. 1997. "Software Reuse: Principles, Patterns, Prospects." *CIT. Journal of Computing and Information Technology* 5 (1): 33–49.

Tari, Zahir. 1992. "Interoperability between Database Models." In *Proceedings of the IFIP WG 2.6 Database Semantics Conference on Interoperable Database Systems (DS-5), Lorne, Victoria, Australia, 16-20 November 1992*, 101–18.

Thüm, Thomas, Sven Apel, Christian Kästner, Ina Schaefer, and Gunter Saake. 2014. "A Classification and Survey of Analysis Strategies for Software Product Lines." *ACM Computing Surveys* 47 (1): 1–45. doi:10.1145/2580950.

May 20th, 2016

## Biography

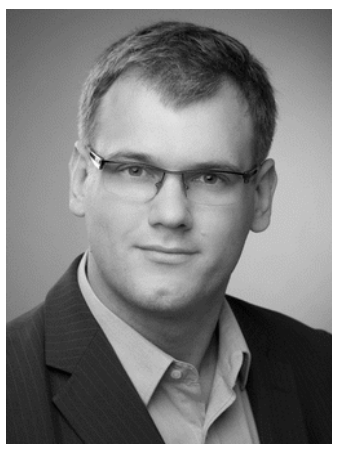

**Aleksander Lodwich** holds a degree in *Master of Science in Computer Vision and Computational Intelligence* (2007) and a *Dipl.-Inform. (FH) in Computerized Communication Systems* (2005) both issued by Fachhochschule Südwestfalen. From 2005 to 2007 he optimized the Vagus Nerve stimulation therapy (OptiVaNeS) with the help of data mining methods. A side goal was to demonstrate reduced amount of necessary animal testing when developing new therapies with data mining. From 2005 to 2009 the author was improving usability of pattern recognition methods for non-experts in the PaREn project at DFKI Kaiserslautern. After parental leave he has joined ITK Engineering AG in Stuttgart in 2012 and has since served as software engineer and software architect for industrial projects and research projects (among them ARTEMIS JU CRYSTAL). His interests are construction of autonomous systems with broad range of capabilities and alternative approaches to engineering and production which would drastically cut cost of engineering and production.

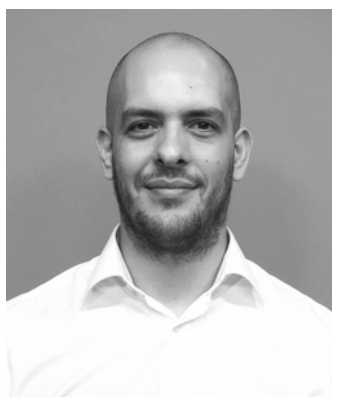

**Dr. Jose María Alvarez Rodríguez** is Master of Computer Science (2007) and Bachelor of Computer Science (2005) by the University of Oviedo. In June 2008 he was rewarded with the "Best Final Degree Project in Computer Science" thanks to this project "Activation of concepts in ontologies through the Spreading Activation Technique". From 2005 to 2010 he worked at the R&D Department in the Semantic Technologies area within CTIC Foundation. He has also participated in more than 18 research projects in different competitive programmes and he is the author of more than 50 publications and other research works. He held a position as part-time Assistant Professor from 2008 to 2012 at the Department of Computer Science within the University of Oviedo. He also worked in the WESO RG (University of Oviedo) and holds a PhD about e -Procurement and Linked Data. In 2012, he was rewarded with a HPC2-Europe Transnational Access Programme grant at SARA (Amsterdam, Netherlands). In 2013, he held a position as Marie Curie Postdoc at SEERC (Thessaloniki, Greece) within the RELATE-ITN FP7 project researching in "Quality Management in Service-based Systems and Cloud Applications". Currently he is Visiting Professor within the Department of Computer Science at the Carlos III University of Madrid. Finally, he is also member of INCOSE and the OSLC Requirements Management working group.